\documentclass[12pt, Arial]{article}
\pdfoutput=1
\usepackage{amsmath,amssymb}
\usepackage{geometry}
\usepackage{pdfpages}
\usepackage{graphicx}
\usepackage{setspace}
\def\ROLLOFF{{\em ROLLOFF }}

\begin{document}
\includepdf[pages={1-21}]{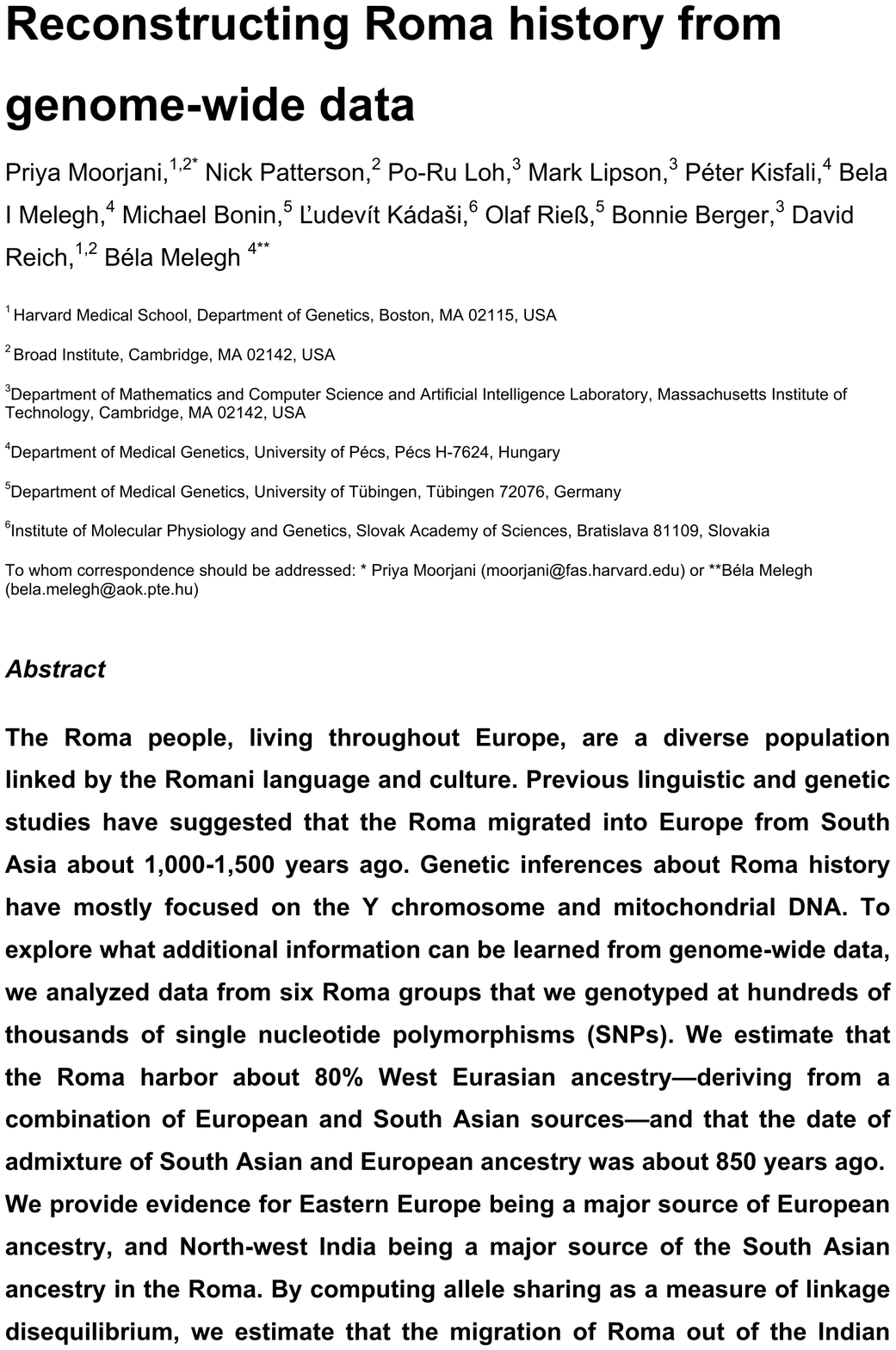}
\includepdf[pages={1-1}]{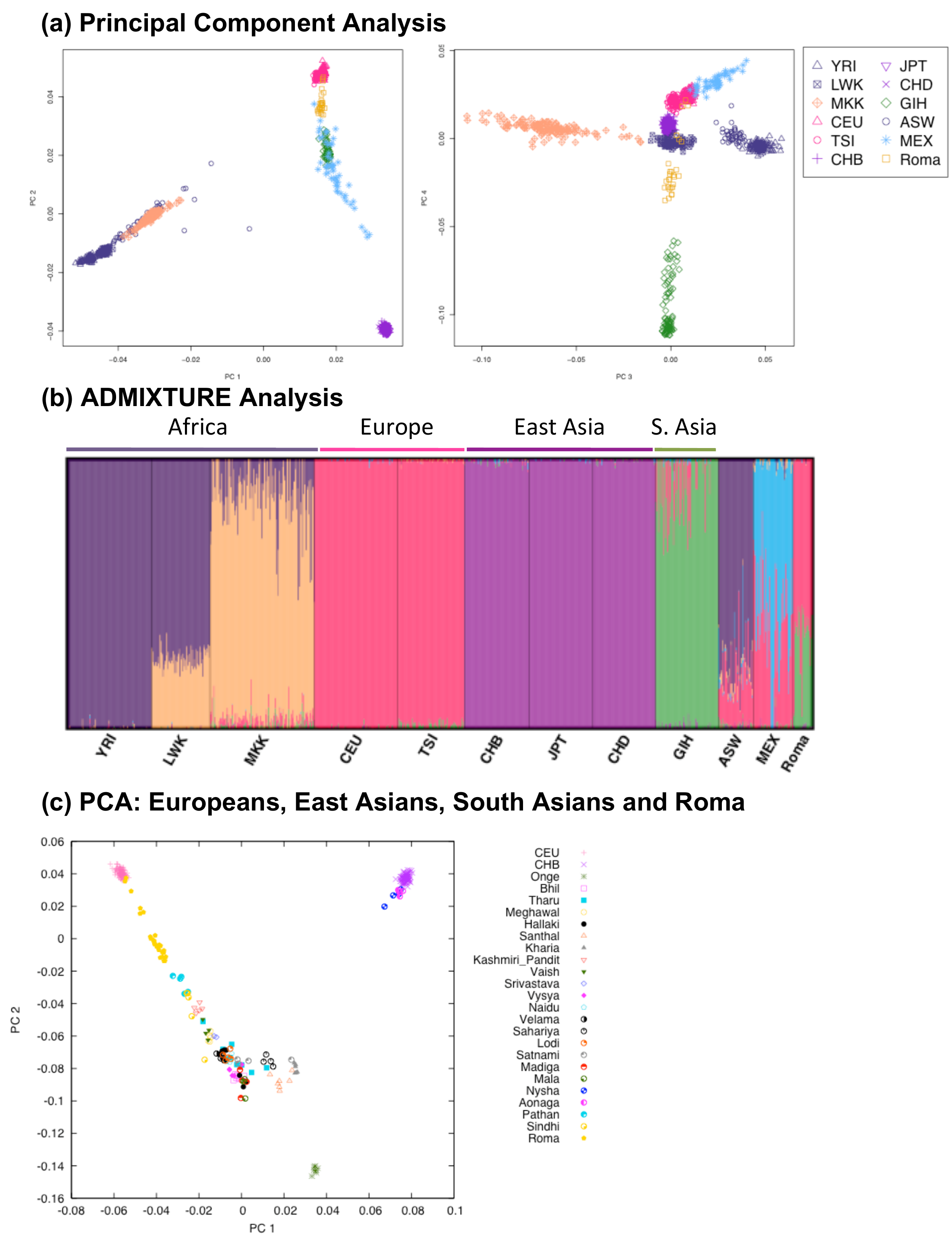}
\includepdf[pages={1-1}]{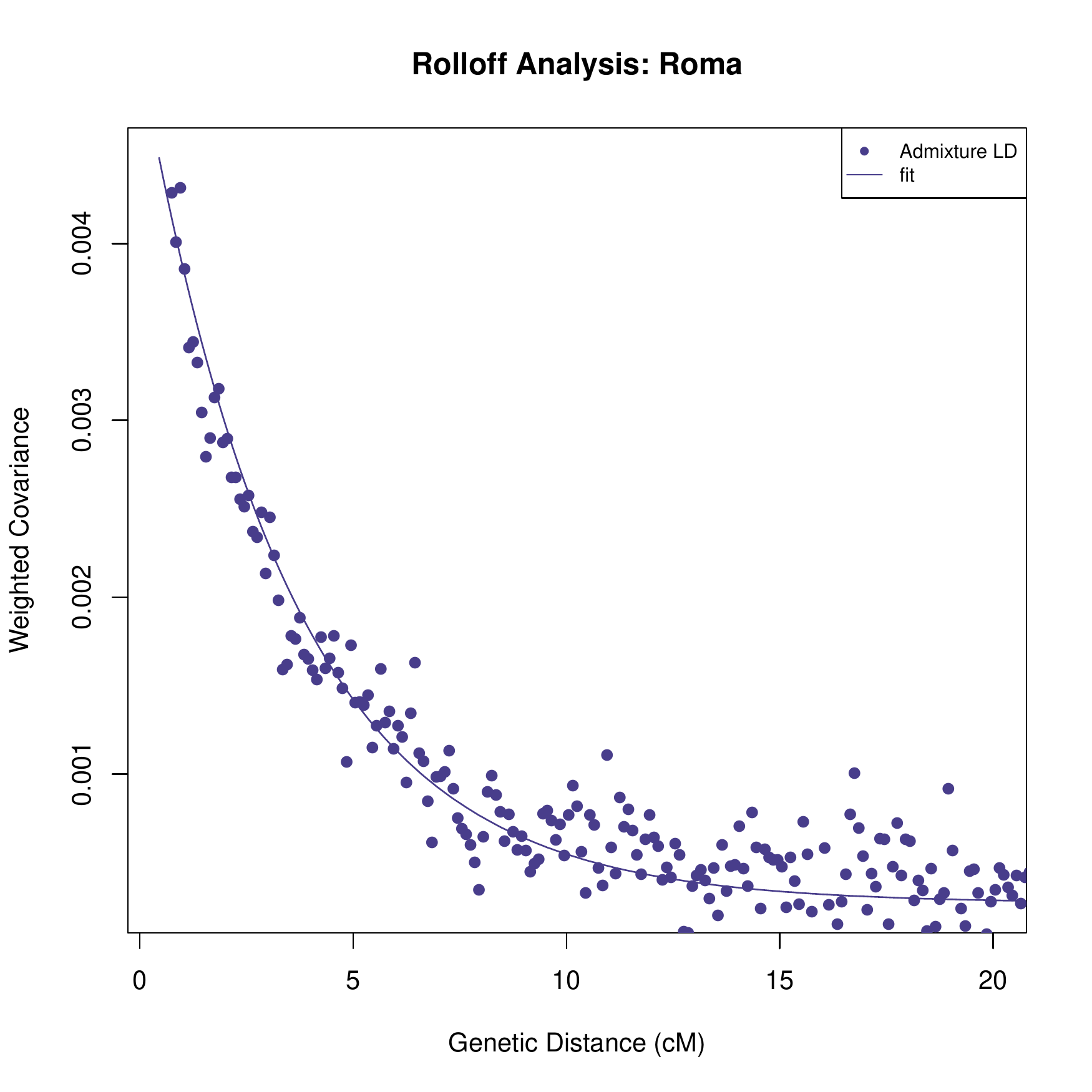}
\includepdf[pages={1-1}]{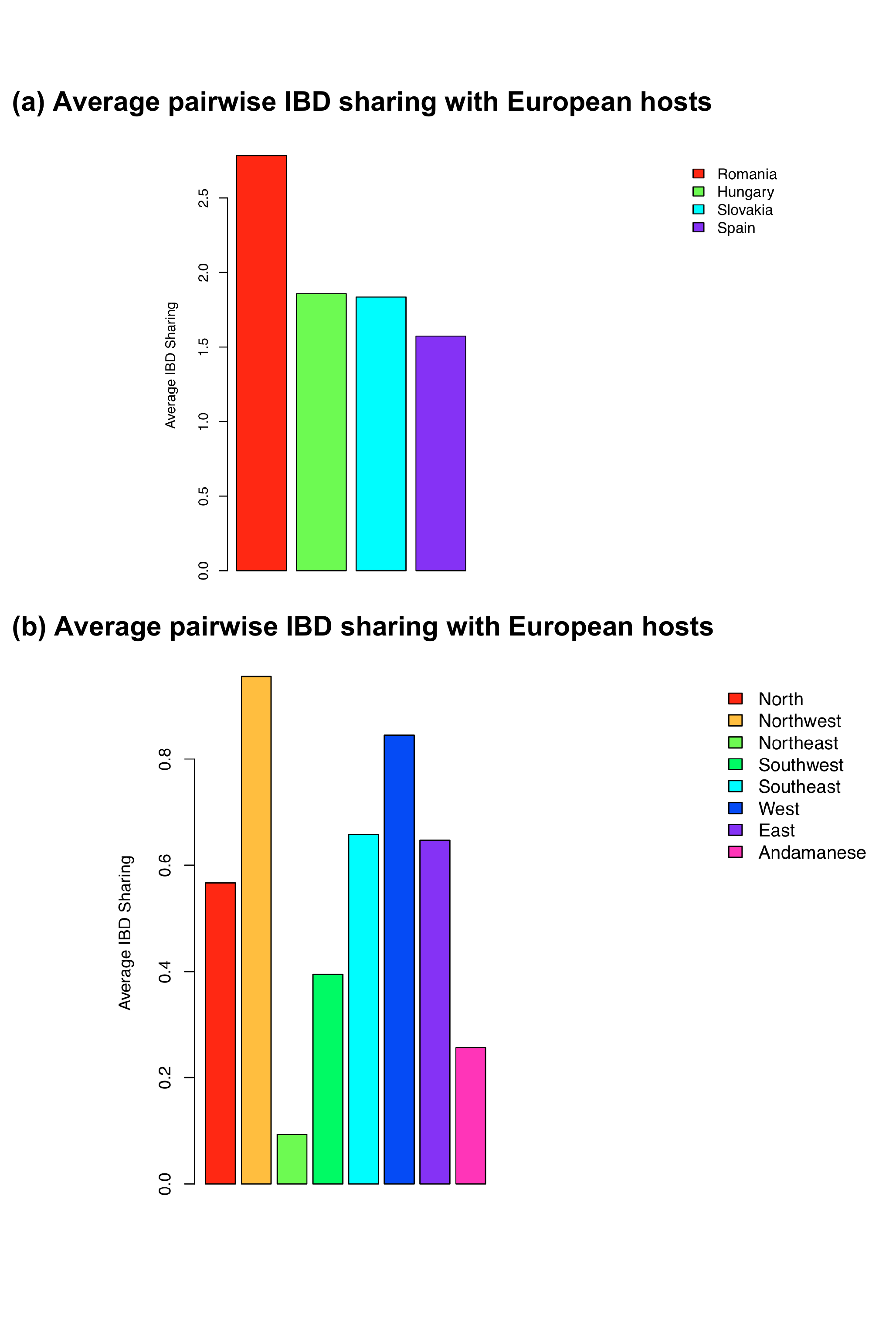}
\includepdf[pages={1-1}]{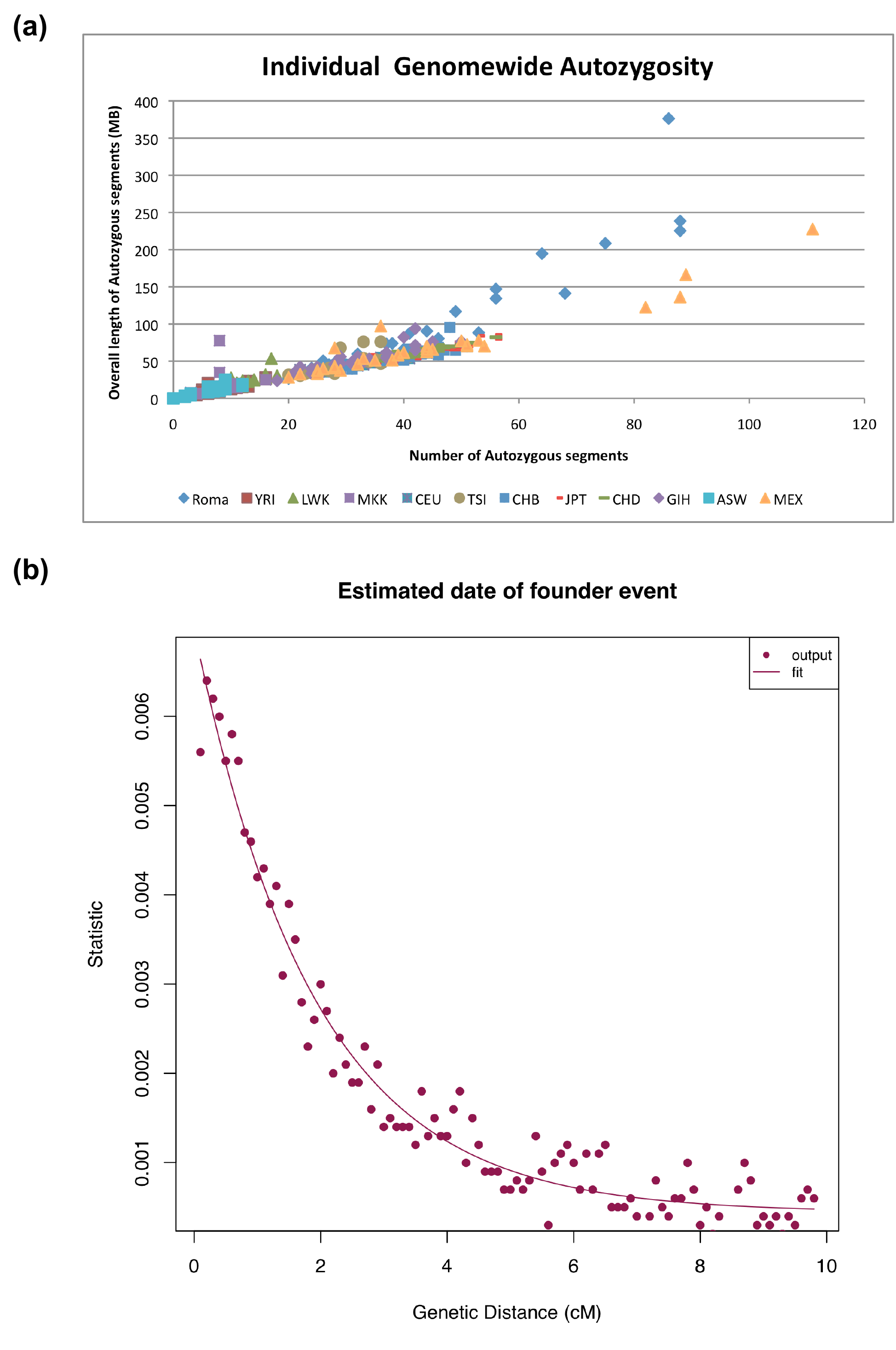}
\includepdf[pages={1-1}]{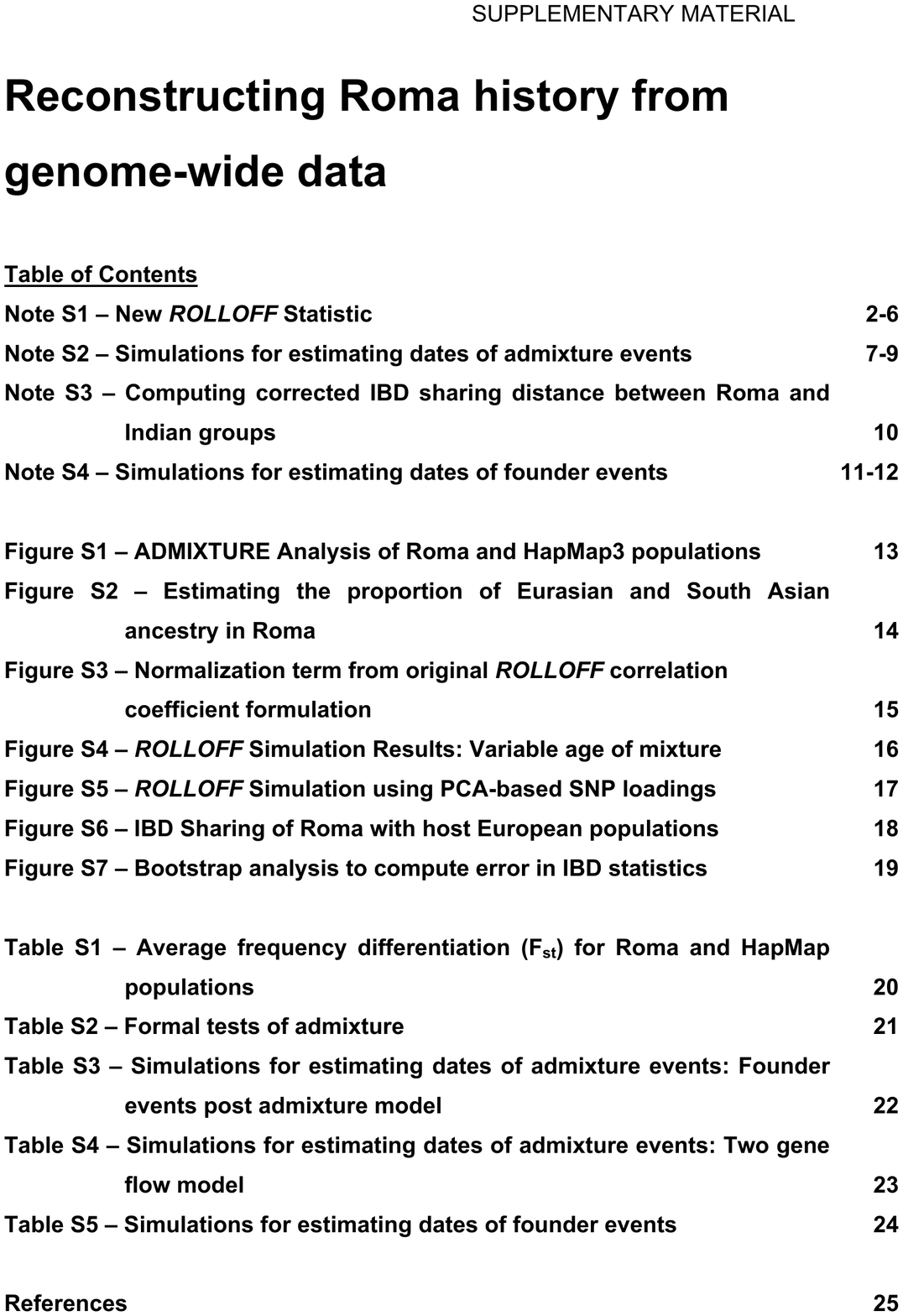}
\setcounter{page}{2}

\subsection*{NOTE S1. New \ROLLOFF statistic}

In this note we consider alternative forms of the \ROLLOFF linkage
disequilbrium (LD) statistic[1] for dating population admixture
events.  We show that the original \ROLLOFF statistic is susceptible to
downward bias in the event of a recent population bottleneck,
and we propose a modification of the statistic that is robust against
such an effect (Table S3).

The \ROLLOFF technique applies two key insights: first, that admixture
creates LD that decays exponentially as recombination
occurs---explicitly, as $e^{-nd}$, where $n$ is the number of
generations since admixture and $d$ is the genetic distance between
SNPs---and second, that the amount of admixture LD between each pair
of SNPs is proportional to the product of the allele frequency
divergences between the ancestral populations at those sites.  The
latter observation allows the $e^{-nd}$ admixture LD decay signal to
be detected (via a SNP-pair weighting scheme) and harnessed to infer
the mixture date $n$.

The original \ROLLOFF statistic captures admixture LD in the form of
SNP autocorrelation.  Defining $z(x,y)$ to be the (Fisher
$z$-transformed) correlation coefficient between SNP calls at sites
$x$ and $y$, \ROLLOFF computes the correlation coefficient between
values of $z(x,y)$ and weights $w(x,y)$ over pairs of SNPs binned by
genetic distance:
\begin{equation} \label{eq:rolloff_original}
A(d) := \frac{\sum_{|x-y| \approx d} z(x,y)w(x,y)}{\sqrt{\sum_{|x-y| \approx d} z(x,y)^2}\sqrt{\sum_{|x-y| \approx d} w(x,y)^2}},
\end{equation}
the idea being that $A(d) \propto e^{-nd}$.

While this setup estimates accurate dates for typical admixture
scenarios, it turns out to be noticeably biased in the case of a
recent bottleneck.  However, we will show that the following
modified statistic does not suffer from the bias:
\begin{equation} \label{eq:rolloff_reg_coeff}
R(d) := \frac{\sum_{|x-y| \approx d} z(x,y)w(x,y)}{\sum_{|x-y| \approx d} w(x,y)^2}.
\end{equation}
(Note that $R(d)$ amounts to taking the regression coefficient of
$z(x,y)$ against the weights $w(x,y)$ for SNP pairs within each bin.)

An additional detail of our \ROLLOFF variant is that we modify
$z(x,y)$ to measure admixture LD as the covariance between SNPs $x$
and $y$ rather than the correlation (i.e., it equals the classical LD
statistic $D$ rather than the correlation $r$).  We believe the use of
covariance rather than correlation for $z(x,y)$ has little impact on
the performance and properties of the statistic (as it roughly amounts
to multiplying by a constant factor) but makes the statistic more amenable
to mathematical analysis.

\subsection*{Explanation of bias from recent bottlenecks}

The bias in the original formulation of \ROLLOFF
(1) introduced by a recent bottleneck
can be readily explained at an intuitive level: the problem is that
while the numerator of the correlation coefficient, $\sum_{|x-y|
  \approx d} z(x,y)w(x,y)$, decays as $e^{-nd}$ as intended, the
normalization term
\begin{equation} \label{eq:z-norm}
\sqrt{\sum_{|x-y| \approx d} z(x,y)^2}
\end{equation}
also exhibits a decay behavior that confounds the $e^{-nd}$ signal
(Figure S3).  The reason is that a strong bottleneck introduces a very
large amount of LD, effectively giving $z(x,y)$ a random large
magnitude immediately post-bottleneck that is independent of the
distance between $x$ and $y$.  This LD subsequently decays as
$e^{-nd}$ until the magnitude of $z(x,y)$ reaches the level of random
sampling noise (arising from the finite sample of admixed individuals
being used to calculate $z$).  In non-bottlenecked cases, the
square-norm of $z(x,y)$ is usually dominated by sampling noise, so the
normalization term (3) effectively amounts to a
constant, and dividing out by it has no effect on the decay rate of
$A(d)$.

The ``regression coefficient'' version of the \ROLLOFF statistic
(2) does not contain the normalization term
(3) and thus does not incur bias from bottlenecks.


\subsection*{Precise effect of genetic drift on original and modified \ROLLOFF statistics}

We now rigorously derive the above intuition.  We will assume in the
following calculations that the \ROLLOFF weights are taken as the
product of allele frequency divergences $\delta(x)$ and $\delta(y)$ in
the ancestral mixing populations:
\[
w(x,y) := \delta(x)\delta(y).
\]
(Our reasoning below applies whether we have the true values of
$\delta(x)$ and $\delta(y)$ or compute weights based on related
reference populations or PCA loadings, however.)  We also assume that
all SNPs are polymorphic ancestrally---i.e., we ignore mutations that
have arisen in the admixed population---and that the SNP ascertainment
is unbiased with respect to the populations under consideration.

For a diploid population of size $N$ with chromosomes indexed by $i =
1, \dots, 2N$, we set
\[
z(x,y) := \frac{1}{2N} \sum_{i=1}^{2N} (X_i-\mu_x)(Y_i-\mu_y)
\]
to be the covariance between binary alleles $X_i$ and $Y_i$ at sites
$x$ and $y$, respectively.  We assume for ease of discussion that the
data are phased; for unphased data, $z(x,y)$ is essentially a noisier
version of the above because of cross terms.

We are primarily interested in the behavior of $z(x,y)$ from one
generation to the next.  Fix a pair of SNPs $x$ and $y$ at distance
$d$ and let $z_0$ denote the value of $z(x,y)$ at a certain point in
time.  After one generation, due to finite population size and
recombination, the covariance becomes[2]
\begin{equation} \label{eq:cov_after_one_gen}
z_1 = z_0e^{-d}(1-1/2N) + \epsilon,
\end{equation}
where $N$ is the population size, $e^{-d}$ is the probability of no
recombination, $(1-1/2N)$ is a Bessel correction, and $\epsilon$ is a noise term with mean 0 and variance on the order of
$1/N$.  Iterating this equation over $n$
generations, the final covariance is
\[
z_n = z_0e^{-nd}e^{-n/2N_e}+\epsilon_{\text{agg}},
\]
where $N_e$ is the effective population size over the interval and
$\epsilon_{\text{agg}}$ is a sum of $n$ partially decayed noise terms.

Now let time 0 denote the time of admixture between two ancestral
populations mixing in proportions $\alpha$ and $\beta := 1-\alpha$.  (The bottleneck may have occurred either before or after this point, as long as it does not influence the calculation of the weights.) 
Then a little algebra shows that
\[
E[z_0] = 2\alpha\beta\delta(x)\delta(y),
\]
assuming the mixture is homogeneous and the distance $d$ is large
enough that background LD can be ignored.  (In practice, heterogeneity
in the admixed population changes the above form and results in the
addition of an affine term to the \ROLLOFF curve which we explicitly
fit.  We also typically fit only data from SNP pairs at distance $d >
0.5 cM$ to avoid background LD.)  We can now compute the modified
\ROLLOFF statistic:
\begin{eqnarray*}
E[R(d)] & = & E\left[\frac{\sum_{|x-y|\approx d}{z(x,y)\delta(x)\delta(y)}}{\sum_{|x-y|\approx d}{\delta(x)^2\delta(y)^2}}\right] \\
& \approx & \frac{\sum_{|x-y|\approx d}[2\alpha\beta\delta(x)\delta(y)e^{-nd}e^{-n/2N_e}+\epsilon_{\text{agg}}]\delta(x)\delta(y)}{\sum_{|x-y|\approx d}{\delta(x)^2\delta(y)^2}} \\
& \approx & 2\alpha\beta e^{-nd}e^{-n/2N_e}.
\end{eqnarray*}
Importantly, in the last step we use the fact that the combined noise term $\epsilon_{\text{agg}}$ is uncorrelated with $\delta(x)\delta(y)$.  Thus,
even a strong bottleneck with a low value of $N_e$ only scales $R(d)$
by the constant factor $e^{-n/2N_e}$, and the $e^{-nd}$ scaling of the
\ROLLOFF curve as a function of $d$ is unaffected.

On the other hand, if we use the original correlation form
\eqref{eq:rolloff_original} of the \ROLLOFF statistic $A(d)$, then the
numerator still has the form of an exponential decay $Ae^{-nd}$, but
now we divide this by the norm $\sqrt{\sum_{|x-y| \approx d}
  z(x,y)^2}$.  In the case of a strong bottleneck, $z(x,y) =
z_0e^{-nd}e^{-n/2N_e}+\epsilon_{\text{agg}}$ can be dominated by the aggregate noise term $\epsilon_{\text{agg}}$.  Indeed, if the bottleneck occurred $k$
generations ago, then the noise terms $\epsilon_i$ from the
time of reduced population size will have decayed by $e^{-kd}$ since
the bottleneck but can still have large variance if the population
size $N_{\text{bot}}$ was very small at the time.  In this case, at
lower values of $d$, $E[z(x,y)^2] =
E[(z_0e^{-nd}e^{-n/2N_e}+\epsilon_{\text{agg}})^2]$ will be dominated by
$E[\epsilon_{\text{agg}}^2]$ which will scale approximately as
$e^{-2kd}/N_{\text{bot}}$.  Hence, the denominator of $A(d)$ will be
significantly larger at low $d$ than at high $d$, causing a partial
cancellation of the exponential decay of the \ROLLOFF curve and thus a
downward bias in the estimated date of admixture.

\includepdf[pages={2-20}]{final-supp.pdf}

\end{document}